\documentstyle[prb,aps]{revtex}
\input epsf
\begin{document}

\preprint{}
\title{Eight-band calculations of strained InAs/GaAs quantum dots
compared with one, four, and six-band approximations.
}
\author{Craig Pryor\cite{email},}
\address{
Department of Solid State Physics \\
Box 118, Lund University\\
S-221 00 Lund, Sweden
}
\maketitle
\begin{abstract}
The electronic structure of pyramidal shaped
InAs/GaAs quantum dots is calculated
using an  eight-band strain dependent $\bf k\cdot p$ Hamiltonian.
The influence of strain on band energies
and the conduction-band effective mass are examined.
Single particle bound-state energies
and exciton binding energies are computed as functions of island size.
The eight-band results are compared with those
for one, four and six bands, and with results from
a one-band approximation in which
$m_{eff}(\vec r)$ is  determined by the local value of
the strain. The eight-band model predicts a lower ground state
energy and a larger number of excited states than the other approximations.
\end{abstract}


\section{Introduction}
Semiconductor quantum dots 
made by Stranski-Krastanow growth have been of great interest
over the past few years.
Such heterostructures are made by epitaxially depositing semiconductor
onto a substrate of lattice mismatched material. The deposited material 
spontaneously forms nm-scale islands which are  subsequently covered
by deposition of the substrate  material. In this way
electrons and holes may be confined within a quantum dot of size
$10 ~\rm nm$ or less.  The islands have a pyramidal shape with
 simple crystal planes for their surfaces. 
The presence of strain significantly
alters the electronic structure of the quantum dot states. 
Theoretical studies of strained islands have employed
various degrees of approximation to the geometry, strain distribution, and
electron dynamics,  ranging from single-band models of 
hydrostatically strained islands, to multiband models including 
realistic shapes and strain distributions. 
\cite{bastard,grundmann,cusack,pryor,fu}

In this paper we consider an InAs island surrounded by GaAs.
Due to the large lattice mismatch  ( $\approx 7 \%$) the strain effects are
substantial. The influence of strain is compounded by the fact that
InAs has a narrow band gap ($E_g = 0.418  ~\rm eV$), implying 
strong coupling between the valence and conduction bands.
This provides a compelling reason to use an eight-band model.
To date, strain-dependent eight-band calculations
have been done for quantum wires\cite{stier}, but not for dots.

\begin{figure}
\epsfxsize=4.0cm
\epsfbox{ 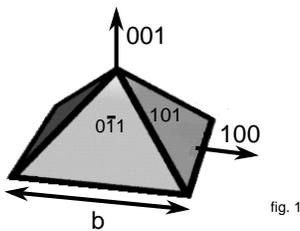 }
\caption{Island geometry.  The island geometry is parameterized 
by the length of the base, b.
\label{fig1}
}
\end{figure}

We assume the island is a simple square-based pyramid with $101$-type planes
for the sides, as shown in Fig. 1. 
The size of the island is parameterized by the length of the base, $b$.
The choice of island shape is somewhat arbitrary. There is no clear
consensus on the exact shape, and 
it may vary
with the details of growth conditions. 
The simple pyramidal geometry of Fig. 1 was chosen primarily
because it has been used in previous calculations,\cite{grundmann,cusack}
hence facilitating comparisons.

An unavoidable consequence of  Stranski-Krastanow island formation is that
 $1-2$ monolayers of island material remains
on the substrate surface. 
This wetting layer is omitted from the calculations because it
 may be accounted for
 separately. The strain is insensitive to the wetting layer primarily
because it is so thin, and also because it is
biaxially strained to match the substrate lattice.
The wetting layer does play a role in the electronic structure
since it provides a quantum well state that is coupled to the quantum dot
state. However, we are most interested in the
 tightly bound quantum dot states. For these states the wetting layer and
quantum dot may be treated separately,
 simplifying the analysis of different wetting layer thicknesses.

After briefly outlining the calculational methods, we examine the 
strain-induced band structure, the 
single particle energies as a function of island size, and the exciton
binding energy. Finally we compare the eight-band results with
calculations using one, four, and six bands.

\section{Calculation}
The technique used to obtain the electronic structure
has been described previously, where is was used for a six-band
calculation of InP islands embedded in GaInP. \cite{pryor}
Here we will focus on the differences due to the use of eight bands.
The entire calculation is done on a cubic grid with
periodic boundary conditions.
First, the strain is calculated using linear  continuum elastic theory.
The strain energy for the system \cite{LL} is computed using a finite
differencing approximation, and then minimized using the conjugate
gradient algorithm. 

The electronic structure is solved in the envelope approximation
using an eight-band  strain-dependent $\bf k\cdot p$ Hamiltonian,
 $H_k+H_s$. \cite{bahder}
The kinetic piece of the Hamiltonian is
\begin{eqnarray}
H_k&=&\left (\matrix{
A & 0 &   V^* & 0 & \sqrt{3}V &- \sqrt{2}U &  -U & \sqrt{2}V^*\cr\cr
0 & A & -\sqrt{2} U & -\sqrt{3}V^*& 0 &  -V & \sqrt{2}V &  U\cr\cr
 V&-\sqrt{2}U&-P+Q    & -S^*   &R     & 0   & \sqrt{3\over2}S  & -\sqrt{2}Q 
            \cr\cr
0& -\sqrt{3} V&-S        & -P-Q       &0     & R  &  -\sqrt{2}R   
 & {1\over\sqrt{2}}S \cr \cr
\sqrt{3}V^*&0&R^*       & 0       & -P-Q    & S^*   & {1\over \sqrt{2}}S^*
 & \sqrt{2}R^*  \cr \cr
-\sqrt{2}U&  -V^*&  0&R^*&S     & -P+Q  & \sqrt{2}Q   & \sqrt{3\over 2}S^*
 \cr \cr
 -U&\sqrt{2}V^*&\sqrt{3\over 2}S^* & -\sqrt{2}R^*       & {1\over \sqrt{2}}S 
 & \sqrt{2} Q  & -P-\Delta  & 0   \cr \cr
\sqrt{2}V& U&-\sqrt{2} Q  &{1\over \sqrt2}S^*& \sqrt{2}R   
 & \sqrt{3\over 2}S & 0 & -P-\Delta     \cr \cr
}\right )   
\end{eqnarray}
where
\begin{eqnarray}
A&=&E_c - {\hbar^2\over 2m_0}(\partial_x^2+\partial_y^2+\partial_z^2),  \cr
P&=&-E_v- \gamma_1{\hbar^2\over 2m_0}(\partial_x^2+\partial_y^2+\partial_z^2),
 \cr
Q&=&-\gamma_2 {\hbar^2 \over 2m_0}(\partial_x^2+\partial_y^2-2\partial_z^2),
 \cr
R&=&\sqrt{ 3} {\hbar^2 \over 2m_0}\left[ \gamma_2(\partial_x^2-\partial_y^2)-
2i\gamma_3 \partial_x \partial_y \right], \cr
S&=& -\sqrt{3} ~\gamma_3 {\hbar^2\over m_0}
 \partial_z(\partial_x-i\partial_y),    \cr
U&=&{-i\over \sqrt{3}}P_0 \partial_z,          \cr
V&=& {-i\over \sqrt{6}}P_0(\partial_x-i\partial_y).
\end{eqnarray}
$P_0$ is the coupling between the conduction and
valence bands, $E_c$ and $E_v$ are the 
(unstrained) conduction and valence-band energies respectively, and $\Delta$
is the spin orbit splitting.
The $\gamma_i$'s are modified Luttinger parameters, defined in terms of the 
usual Luttinger parameters, $\gamma_i^L$,  by 
\begin{eqnarray}
    \gamma_1&=&\gamma_1^L-{E_p\over 3E_g+\Delta}\cr
    \gamma_2&=&\gamma_2^L-{1\over 2}{E_p\over 3E_g+\Delta}\cr
    \gamma_3&=&\gamma_3^L-{1\over 2}{E_p\over 3E_g+\Delta}
\end{eqnarray}
where $E_g = E_c - E_v$ is the energy gap, and $E_p = 2m_0 P_0^2 /\hbar^2$.

The strain enters through a matrix-valued potential that couples the various 
components,
\begin{eqnarray}
H_s&=&\left (\matrix{
a_c e & 0 & -v^* & 0 & -\sqrt{3}v & \sqrt{ 2 }u &   u & -\sqrt{2}v^*\cr\cr
0 & a_c e & \sqrt{2} u & \sqrt{3}v^*& 0 & v& -\sqrt{2}v &  -u \cr\cr
-v & \sqrt{2}u&-p+q   & -s^*    &r & 0 & \sqrt{3\over2}s & -\sqrt{2}q 
    \cr\cr
0 & \sqrt{3}v &-s & -p-q &0  & r  &  -\sqrt{2}r  & {1\over\sqrt{2}}s  \cr \cr
-\sqrt{3}v^* & 0 &r^*  & 0  & -p-q  & s^*  & {1\over \sqrt{2}}s^* 
& \sqrt{2}r^*             \cr \cr
  \sqrt{2} u& v^* &0  & r^*  &s  & -p+q  & \sqrt{2}q  & \sqrt{3\over 2}s^* 
   \cr \cr
 u & -\sqrt{2}v^*  &   \sqrt{3\over 2}s^* & -\sqrt{2}r^*       & {1\over 
\sqrt{2}}s  & \sqrt{2} q  & -a_ve  & 0 \cr \cr
 -\sqrt{2}v &   -u & -\sqrt{2} q              &{1\over \sqrt2}s^* & \sqrt{2}r
  & \sqrt{3\over 2}s & 0  & -a_ve  \cr \cr
} \right )   \label{Hs}
\end{eqnarray}
where
\begin{eqnarray}
p&=&a_v(e_{xx}+e_{yy}+e_{zz}), \cr
q&=&b ~[ e_{zz}-{1\over 2} (e_{xx}+e_{yy} ) ],  \cr
r&=&{\sqrt{3}\over 2}~b~(e_{xx}-e_{yy}) -ide_{xy},   \cr
s&=&-d (e_{xz} - ie_{yz}  ),  \cr
u&=& {-i\over \sqrt{3}}P_0\sum_{j}{  e_{zj}\partial_j  },   \cr
v&=&{-i\over \sqrt{6}}P_0 \sum_{j}{  ( e_{xj}-ie_{yj} }  )\partial_j.  
\end{eqnarray}
$e_{ij}$ is the strain tensor,
 $b$ and $d$ are the shear deformation potentials, 
 $a_v$ is the hydrostatic valence-band deformation potential, and
 $a_c$ is the conduction-band deformation potential.

In addition to the explicit strain dependence in $H_s$,
there is a small piezoelectric effect which is included.
The strain induced polarization of the material contributes an additional
electrostatic potential which  breaks 
 the $C_4$ symmetry of the islands to  $C_2$. \cite{grundmann}

The energies and wave functions are computed by replacing
 derivatives  with
differences on the same cubic grid used for the strain calculation.
The material parameters and strain in Eq.'s 1-5 vary from site to site.
The Hamiltonian is then a sparse matrix  which is easily
diagonalized using the Lanczos algorithm.
 The calculation is further
simplified by eliminating unnecessary barrier material, since
bound states fall off exponentially within the barrier.

\section{Material Parameters}

The values used for the various material parameters are given in Table I.
All the parameters were set to the values corresponding to the local 
composition, except for the dielectric constant, $\epsilon_R$, which was
set to the value for InAs throughout the structure.
Most parameters were  taken from direct  measurements, 
however a few merit comment.
Neither 
$a_c$ nor $a_v$  have been directly measured,
although the gap deformation potential $a_g = a_v + a_c$ 
has been measured. \cite{LB}  Using the fact that for most  III-IV
semiconductors   $a_c/a_v \approx 0.1$, \cite{DP}  $a_c$ and $a_g$ can be
estimated.
Another important parameter is
the unstrained valence-band offset, $E_{vbo}$ defined as 
$E_v({\rm InAs}) -E_v({\rm GaAs}) $ in the absence of strain. The
value used is based on transition-metal impurity spectra, and is in agreement
with the value from Au Schottky barrier data. \cite{offset}
The value is derived using the fact that transition metal
impurities are empirically observed to
have energy levels fixed with respect to the vacuum, relatively independent
of their host environment.
Thus, by comparing band edges referenced
to the impurity levels in two different materials one deduces the relative
band offsets if the strain could be turned off.
The  ground state energies of Mn impurities are  
$0.028 ~\rm eV$  and $0.113 ~\rm eV$
above the valence band in
InAs and GaAs respectively \cite{LB}, so the InAs valence band is 
$85 ~\rm meV$ above  GaAs.
\vbox{
\begin{table}
\caption{Material parameters. Unless otherwise noted,  values are taken
         from Landolt-B\"{o}rnstein. $e_{14}$ is the piezoelectric constant,
         $\epsilon_R$ is the relative dielectric constant, the $C$'s are
         the elastic constants, and $a$ is the lattice constant.
                     \label{table1}
                }
\begin{tabular}{lcr}
Parameter&InAs&GaAs\\
\tableline
$\gamma^L_1$ & $19.67$                               &  $6.85$          \\
$\gamma^L_2$ & $8.37$                                & $2.1$             \\
$\gamma^L_3$ & $9.29$                                & $2.9$              \\
$E_g$        & $0.418 ~\rm eV$                       &    $1.519 ~\rm eV$ \\
$\Delta$     & $0.38 ~\rm eV$                        & $0.33 ~\rm eV$       \\
$E_p$        & $22.2~\rm eV$                         & $25.7 ~\rm eV$        \\
$a_g$        & $-6.0~\rm eV$    & $-8.6 ~\rm eV$~\tablenotemark[1]         \\
$a_c$        & $-6.66~\rm eV$   & $-9.3 ~\rm eV$~\tablenotemark[1]          \\
$a_v$        & $0.66 ~\rm eV$   & $0.7 ~\rm eV$ ~\tablenotemark[1]           \\
$b$          & $-1.8 ~\rm eV$   &  $-2.0 ~\rm eV$           \\
$d$          & $-3.6 ~\rm eV$   & $-5.4 ~\rm eV$             \\    
$e_{14}$     & $0.045 ~\rm  C/m^2$~\tablenotemark[2] & $0.159 ~\rm 
                      C/m^2$~\tablenotemark[2]         \\
$\epsilon_R$ & $15.15$          & $15.15$~\tablenotemark[3]    \\
$C_{xxxx}$   & $8.329 \times 10^{11} ~\rm dyne/cm^2$ 
             & $12.11 \times 10^{11} ~\rm dyne/cm^2$   \\
$C_{xxyy}$   & $4.526 \times 10^{11} ~\rm dyne/cm^2$ 
             & $5.48 \times 10^{11} ~\rm dyne/cm^2$    \\
$C_{xyxy}$   & $3.959\times 10^{11} ~\rm dyne/cm^2$  
             & $6.04 \times 10^{11} ~\rm dyne/cm^2$        \\
$a$          & $0.60583 ~\rm nm$                     
             & $0.56532 ~\rm nm$                          \\
$E_{vbo}$    & $85~\rm meV$~\tablenotemark[4]                          & - \\
\end{tabular}
\tablenotetext[1]{ Reference \cite{DP}}
\tablenotetext[2]{ Reference \cite{e14} }
\tablenotetext[3]{ Value for GaAs used.}
\tablenotetext[4]{ see text.}
\end{table}
}

\section{Band  Structure}

Some insight may be obtained by examining the strain induced
modification to the band structure.  
Fig. 2 shows the band energies computed
using the local value of the strain (i.e. the eigenvalues of
Eq.~\ref{Hs} with $\vec k = 0$). 
Since the coupling between
conduction and valence bands is proportional to $\vec k$,
for $\vec k=0$ the model reduces to a six-band model with a
decoupled conduction band.
The bands are shown for an island with $b=10~\rm nm$. 
The band diagrams for a 
different sized island are obtained by simply rescaling the x-axes.
The dominant effect of the strain is that the island
experiences a large increase in its  band-gap due to the considerable 
hydrostatic pressure.
The conduction band still has a
potential well $0.4~\rm eV$ deep at the base of the island,
tapering  to $0.27~\rm eV$ at the tip.

The valence band has a  more complex structure. If we could somehow turn off
the strain,
the holes would be confined to the InAs by only $E_{vbo}=85~\rm meV$.
Strain  alters this considerably, and makes the dominant contribution
to the hole confinement.
The most notable features are that  the valence band is peaked near the
tip of the island, with another high point near the base and
a band crossing in between.
This is most
clearly seen in the plot of band energies along the $001$ direction.
The plot along the  $100$  direction near the base  (Fig. 2b) shows 
that there is a slight peak in the valence band at the edge of the base,
a feature shared with InP/GaInP islands.
InP/GaInP islands also have such peaks in the valence band, and they
are sufficiently strong that holes are  localized near
the peaks rather than spread out over the whole island.\cite{pryor}
In InP/GaInP islands there is also a valence band peak in the
barrier material above the island which provides a separate pocket
in which holes may be confined.
From Fig. 2b we see that InAs islands have an elevation of the valence band
above the island, but inside the island the valence-band edge is even higher.
Hence we do not expect holes to be trapped in the barrier material.

\begin{figure}
 \hbox{
  \epsfxsize=8.75cm
  \epsfbox{ 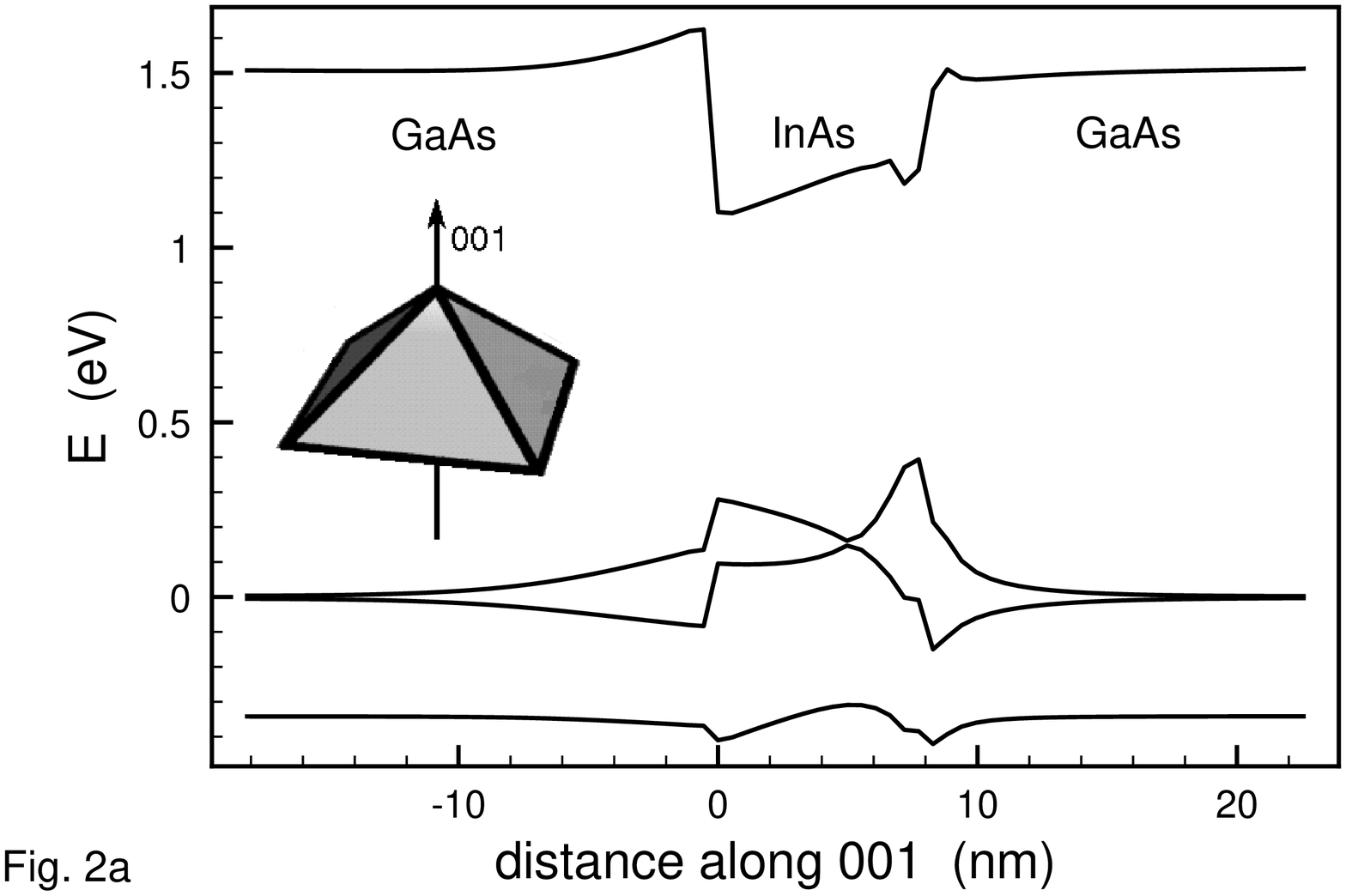 }
  \epsfxsize=8.75cm
  \epsfbox{ 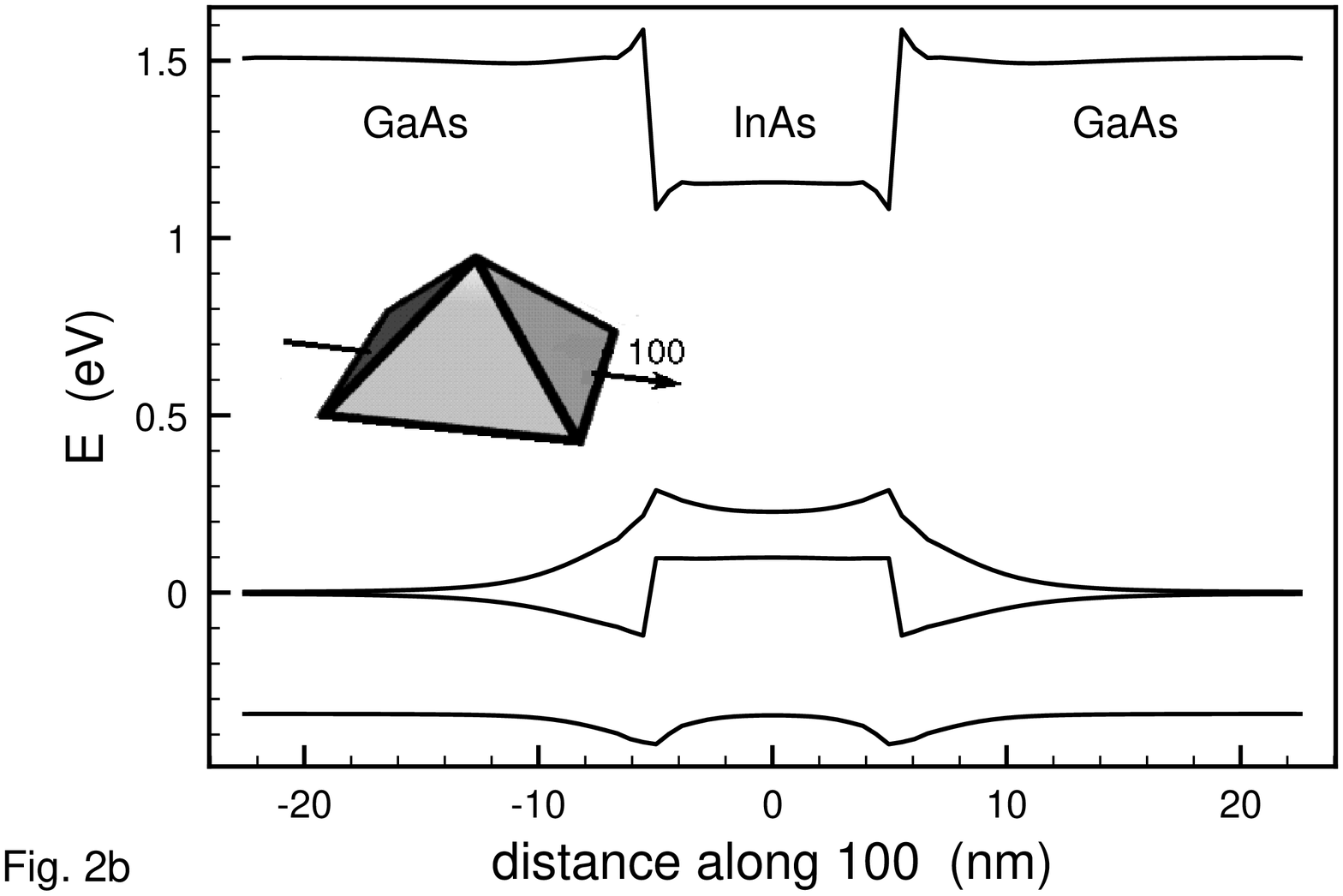 }
 }
 \caption{Band structure based on the local value of  the strain.
 (a) Bands along the $001$ direction, through the center of the island. 
 (b) Bands Along the $100$ direction, through the base   of the island.
 }
 \label{fig2}
\end{figure}

\begin{figure}
  \epsfxsize=8.75cm
  \epsfbox{ 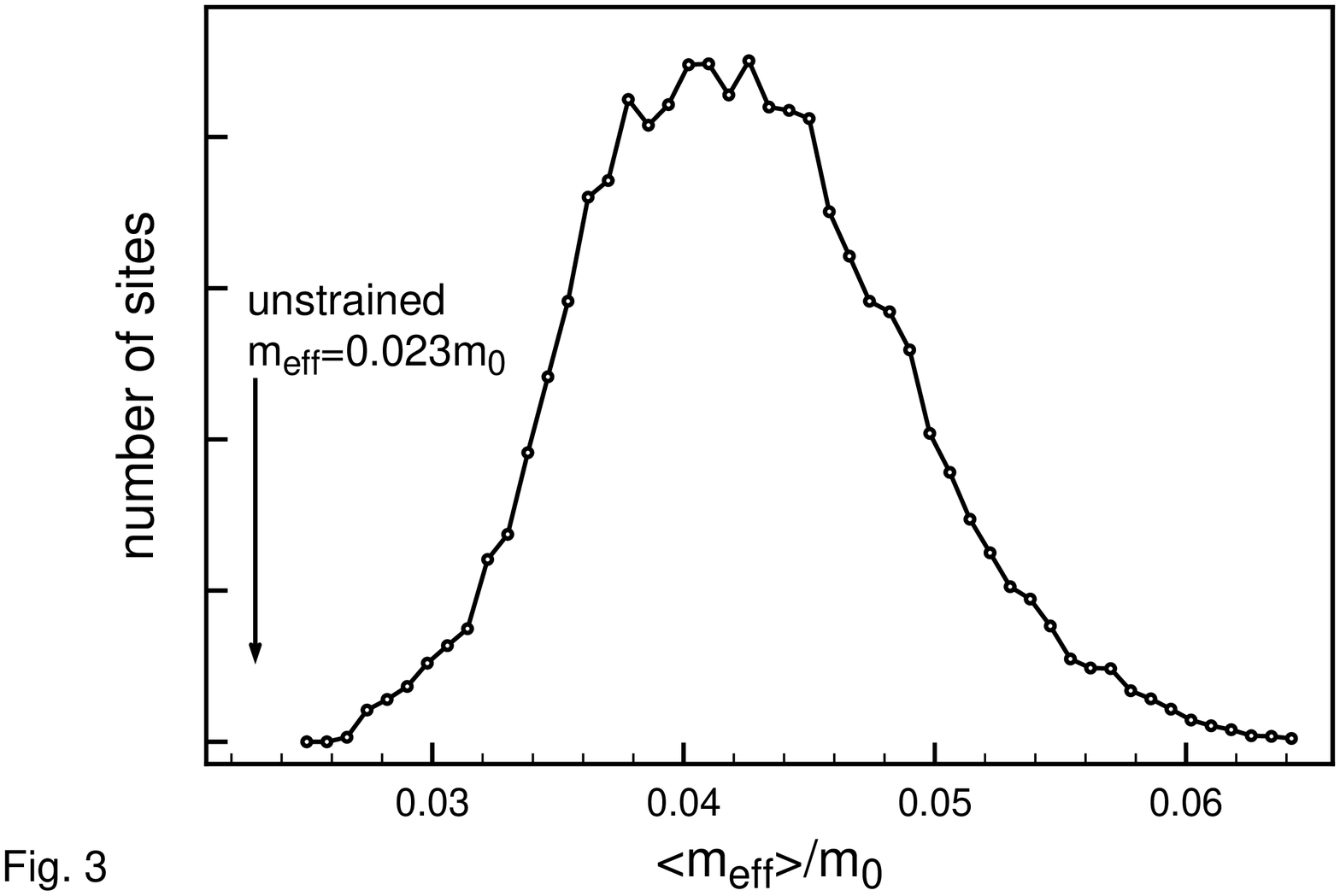 }
  \caption{
      Histogram of the conduction-band effective mass within the island.
      The effective mass is averaged over direction.
  }
\label{fig3}
\end{figure}

Strain-dependent effective masses may be found
by computing the dispersion
relation using the local value of the strain. Since $m_{eff}$ is anisotropic,
it is necessary to average over directions. The valence-band
anisotropies are sufficiently strong that a hole effective mass is of
dubious value.
The conduction-band anisotropy is considerably smaller, however, making
an electron isotropic effective mass a reasonable approximation.
Fig. 3 shows a histogram of $m_{eff}$ within the island.
Due to the large hydrostatic strain, $m_{eff}$ is doubled throughout
much of the island, although there is considerable variation.

\section{Bound States}
The bound state energies were 
computed as a function of island size using the full eight-band 
Hamiltonian (Fig. 4).  Because the calculations were performed assuming no 
wetting layer there are states right up to the GaAs band edges.
The energies for one and two-monolayer wetting layers are also
shown in Fig. 4 for comparison. These were calculated as independent
quantum wells using the envelope approximation.
It should be noted that there will
be a bound quantum dot state regardless of island size.
It is a well known fact that
in one and two dimensions an arbitrarily weak attractive potential
has at least one bound state.
In three dimensions there is no such guarantee, and hence it is
possible to construct a quantum dot which has no bound states.
However the wetting layer forms a quasi 
two-dimensional system with the island acting as an attractive potential.
Therefore, we expect at least one wetting layer state to be bound
to the dot, no matter how small it is.

\begin{figure}
 \hbox{
  \epsfxsize=8.75cm
  \epsfbox{ 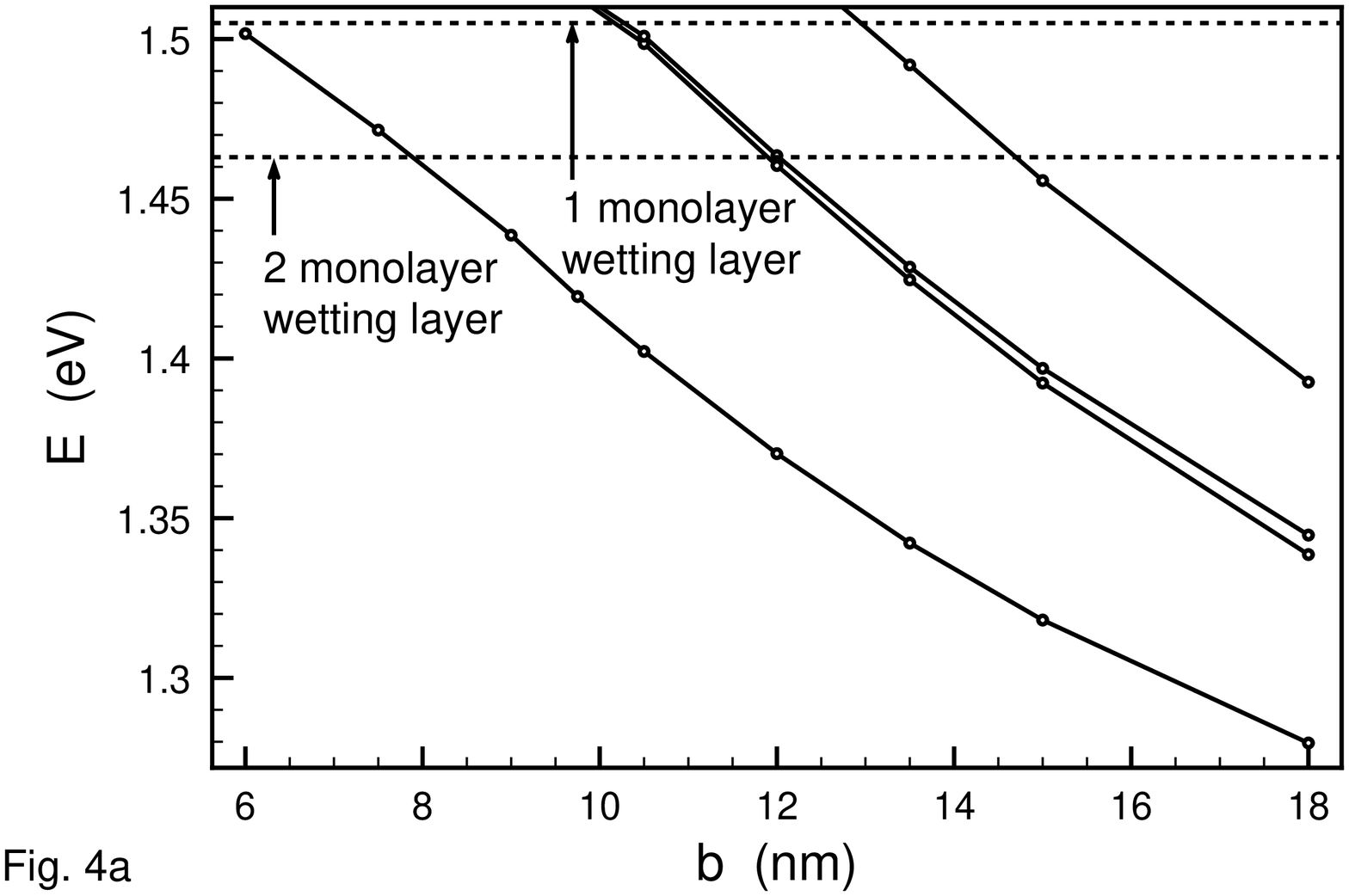 }
  \epsfxsize=8.75cm
  \epsfbox{ 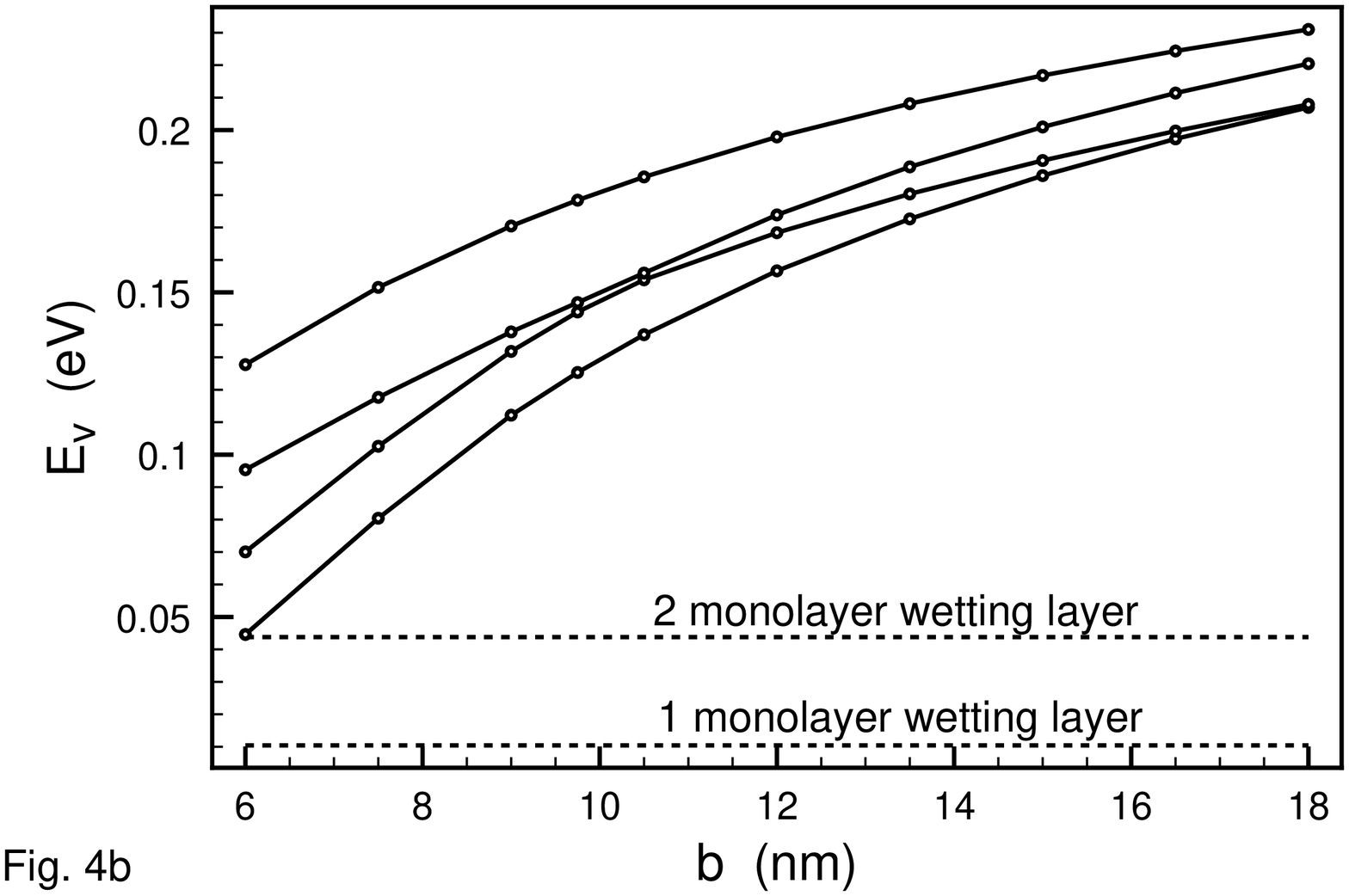 }
 }
 \caption{ Bound state energies as a function of island size.
   The dotted line indicates the energy
   for a one-monolayer biaxially strained InAs wetting layer,
   also computed  in the envelope approximation.
   (a) Conduction band.
   (b)   Valence band.
  }
\label{fig4}
\end{figure}

For the conduction-band there are only a few bound states in the island, as
shown in Fig. 4a. 
The number of excited states in the quantum dot depends on the island's size
and the wetting layer thickness.
In order to have an excited conduction-band
state requires $b>10 ~\rm nm$ and $b>12 ~\rm nm$ for
one and two monolayer wetting layers respectively.
The first excited state is accompanied by a nearly degenerate state.
The splitting between these two states varies from
$2~\rm meV$ to  $6~\rm meV$   for $10 {~\rm nm} < b  < 18 {~\rm nm} $.
The near degeneracy reflects the $C_4$
symmetry of the square island, with the splitting due to the piezoelectric
effect. A  third excited state appears for $b>13 ~\rm nm$ and $b>14~\rm nm$
for one and two monolayers respectively. These limits on $b$ are all upper
bounds since the actual quantum dot energies will be lowered by the coupling
to the wetting layer.
In addition to the change in energy with size, the spacings
vary as well. The gap between the ground state and
first excited state varies from  $60 ~\rm meV$ to $95 ~\rm meV$ over the
range $10 ~{\rm nm} < b < 18 ~{\rm nm} $.

The valence-band states are more strongly confined, due to 
their larger effective mass. Only the first four states are shown in  Fig. 4b,
all of which lie well above the wetting layer energy.
The energy spacings vary from a few meV to $30 ~\rm meV$ over the range
of island sizes considered.

When the island is occupied by an electron and a hole, there will
be additional binding energy from the coulomb interaction.
Ground state exciton energies were computed in the Hartree
approximation using eight-band solutions for both electrons and holes. 
That is, the exciton wave function was assumed to be of the form
$\Psi_{ij}(\vec r_e, \vec r_h)=\psi^e_i(\vec r_e) \psi^h_j(\vec r_h)$.
 $\psi^e$ and $\psi^h$  were found by self-consistent iteration,
with convergence to within $0.1~\rm meV$ usually taking only two  iterations.
Fig. 5 shows the exciton binding energy as a function of island size.
The results are in good agreement with single-band calculations
\cite{grundmann}, which is not surprising
since the coulomb energy  depends only on the charge density for the electron
and hole parts of the wave function.
The single-particle electronic Hamiltonian only affects the coulomb
energy insofar as it alters the charge distribution.
The exciton binding energy increases with decreasing island size,
reaching $27 ~\rm meV$ for $b=9 ~\rm nm$.
Fig. 6 shows the exciton wave function for $b=14 {~\rm nm} $. In spite
of the complex band structure seen in Fig. 2, the electron and hole
wave functions appear ordinary.
The wave functions are spread out over most of the
island, with no signs of localization around smaller regions.
\begin{figure}
  \epsfxsize=8.75cm
  \epsfbox{ 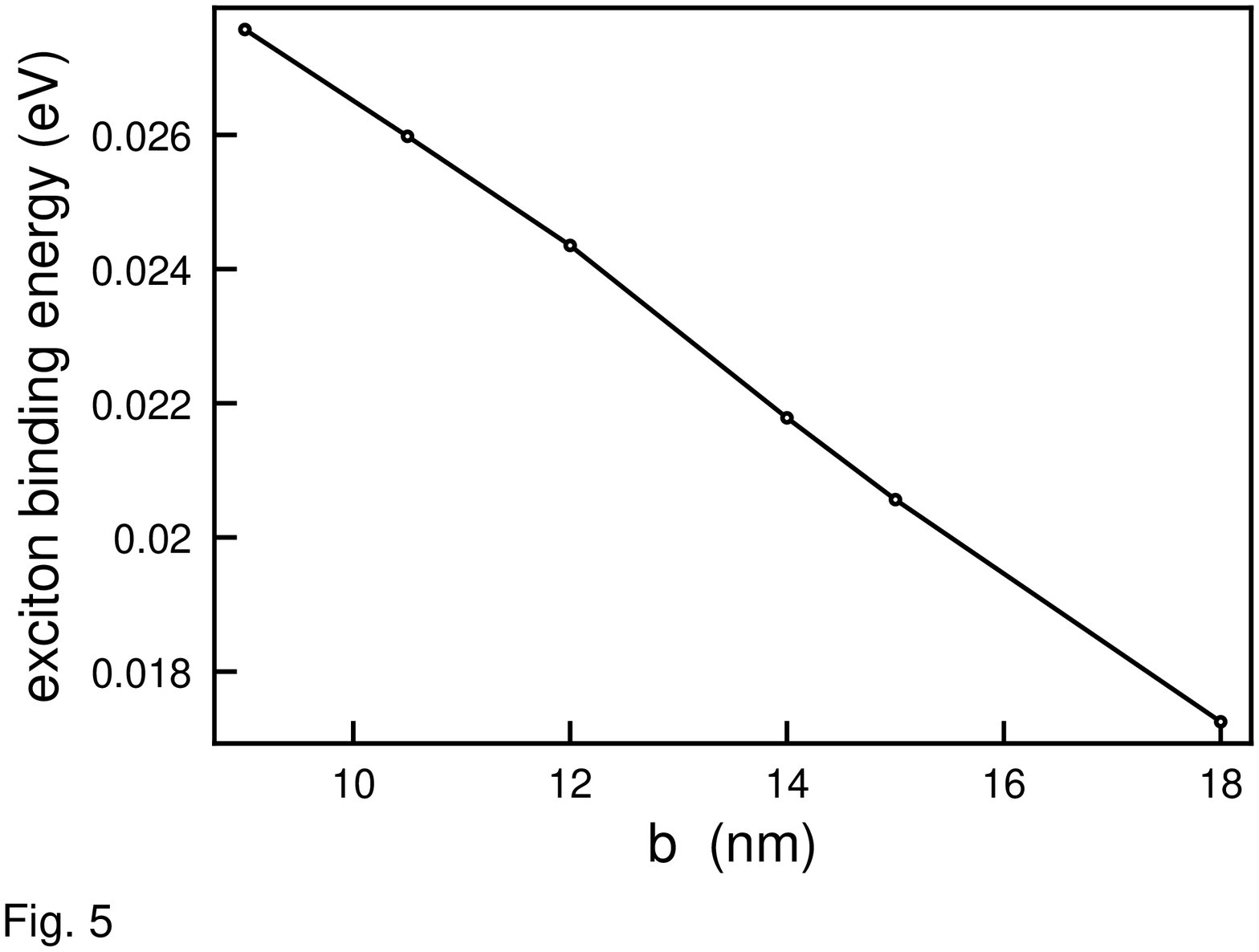 }
  \caption{
    Exciton binding energy versus island size. Exciton binding energies 
    were computed  in the Hartree approximation.
  }
\label{fig5}
\end{figure}

\begin{figure}
  \epsfxsize=8.75cm
  \epsfbox{ 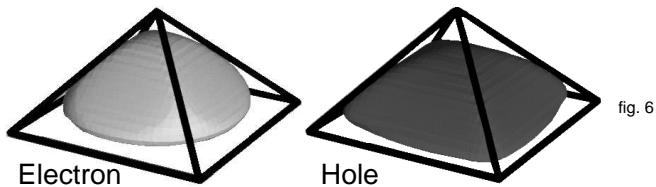 }
  \caption{
       Electron and hole wave functions for the ground state exciton in the  
       Hartree approximation, with $b=14 ~\rm nm$.
       Surfaces are $\sum_{i=1}^{8}{}  |\psi_i(\vec r ) |^2$ equal to
       $0.1 $ of the peak value.
  }
\label{fig6}
\end{figure}

\section{Comparison of Approximations}
There remains the question of whether or not 
an eight-band model is worth the trouble.
Previous authors have used various approximations to reduce
the number of bands.
Grundmann {\it et. al. }
 \cite{grundmann}  treated the electrons and heavy holes as single particles
moving in the strain induced potential corresponding to the band edges.
The effective masses were different in the island and barrier materials, but
took constant unstrained  values within each region.
As was pointed out by Cusack { \it et. al.}
 \cite{cusack}  the narrow InAs band gap
  leads to significant band mixing,
resulting in large strain induced shifts in the effective mass.
Based on a pseudopotential calculation, Cusack {\it et. al.} set
$m_{eff}=0.04 m_0$ for the conduction band,
which is the value predicted
for bulk InAs under the average 
hydrostatic strain in the island.
The valence-band states were calculated using a four-band
model. Note that $m_{eff}=0.04 ~m_0$ is in good agreement with the peak
of the distribution for $m_{eff}$  shown in Fig. 3.

Unfortunately, a comparison with previous calculations
is complicated by the fact that the methods have differed in
more than the Hamiltonian.
Different material parameters were
used, the strain was calculated differently 
(continuum elasticity \cite{grundmann}
versus valence force field method \cite{cusack}),
 and different numerical
techniques were used to solve Schr\"{o}dinger's equation (real space
differencing \cite{grundmann}  versus a plane wave basis \cite{cusack}).
To more directly compare these different approximations,
energies were calculated using several
different Hamiltonians, but all using the same grid and strain profile.
 For the conduction band the methods considered are
({\it i}) setting $m_{eff}$ set to its unstrained values of $0.023 m_0$ in
    the InAs, and $0.0665 m_0$ in the GaAs,
({\it ii}) using the value corresponding to the average
    hydrostatic strain in the InAs $m_{eff}=0.04 m_0$, and using
    $m_{eff}=0.0665 m_0$ in the GaAs,
({\it iii})  using a spatially-varying strain-dependent $m_{eff}(\vec r)$,
({\it iv}) using full eight-band Hamiltonian.

A comparison of the different conduction-band energies
for $b= 14 ~\rm nm$  is
 shown in Fig. 7a. The dominant
 feature  is that the energies decreases as
 the Hamiltonian includes more physics. For the simple unstrained
effective mass only a single state is found. With $m_{eff}=0.04 m_0$
the  binding energy of the
 ground state increases by $ 30 ~\rm meV$. Also, the nearly
degenerate first and second excited states are brought below the energy
of a one-monolayer wetting layer.
Using a one-band model with $m_{eff}(\vec r)$
gives energies very close to those for $m_{eff}=0.04 m_0$.
The eight-band results are significantly different. Not only is the
ground state lower by another $27 ~\rm meV$, but the two nearly degenerate
excited states are clearly confined. In addition, a
third excited state falls below the the one-monolayer energy.
The one-band models with $m_{eff}=0.04 m_0$ and $m_{eff}=m_{eff}(\vec r)$
both predict $E_1 - E_0 \approx 110~\rm meV$. The eight-band model predicts
$E_1 - E_0 \approx 80 ~\rm meV$.

\begin{figure}
  \hbox{
    \epsfxsize=8.75cm
    \epsfbox{ 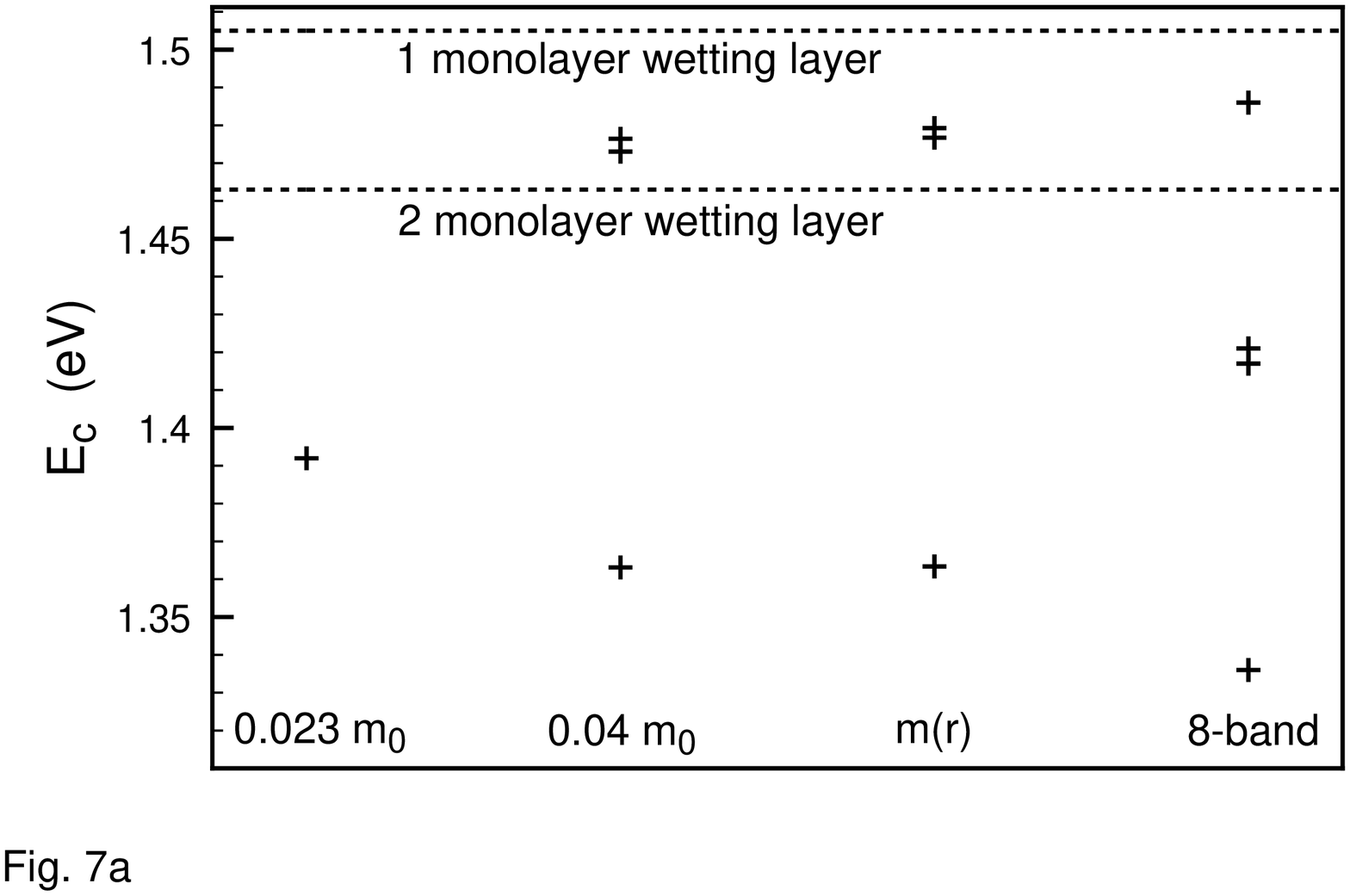 }
    \epsfxsize=8.75cm
    \epsfbox{ 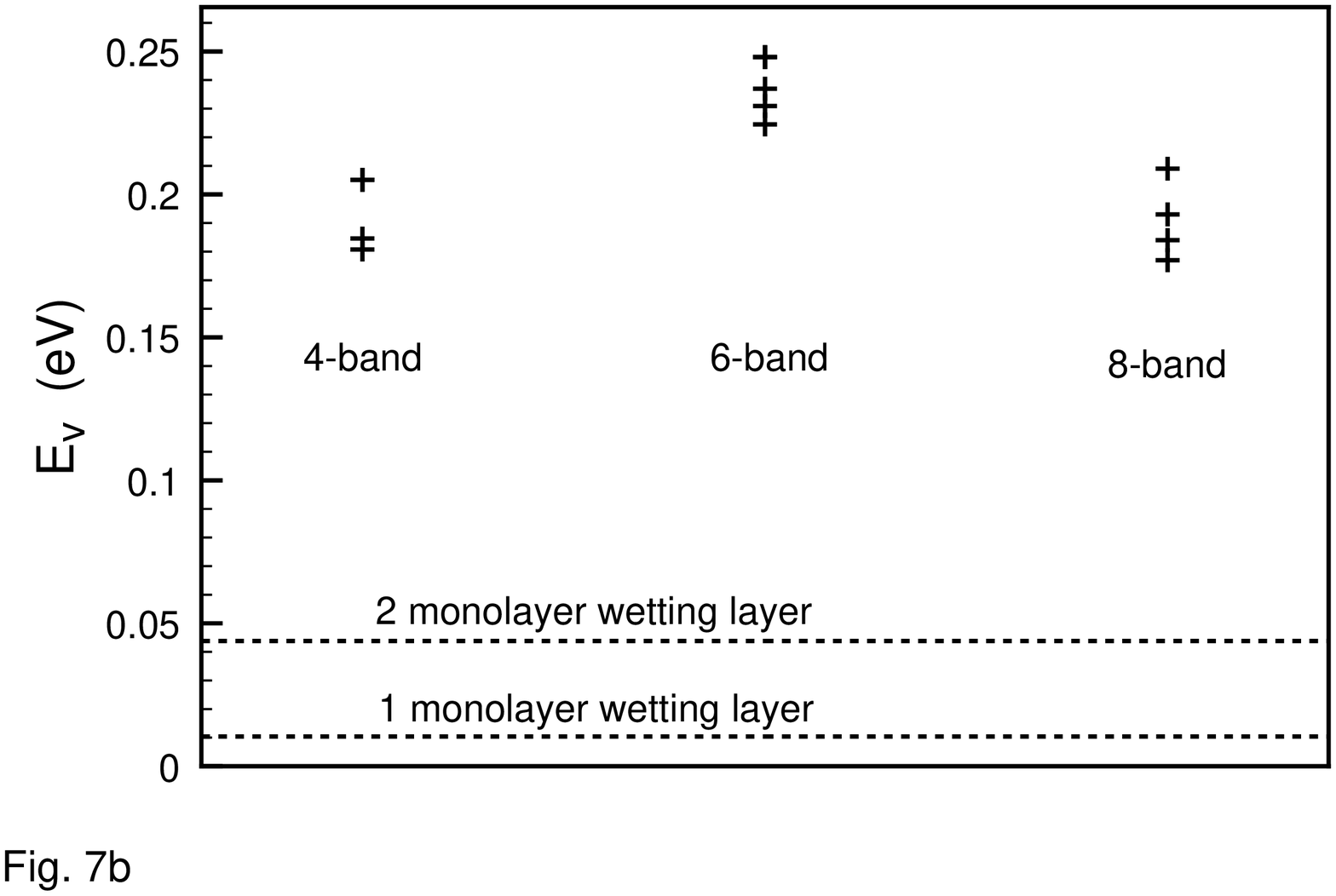 }
  }
  \caption{
	Confined state energies for an island with $b={14~\rm nm} $
	(a) Conduction band
	computed using constant masses, spatially varying mass, and
	eight-band Hamiltonian.
	(b) Valence-band states computed using four-, six-, and eight-band
	Hamiltonians.
  }
\label{fig7}
\end{figure}

The valence-band energies were calculated using four, six and eight-band 
models (Fig. 7b).  The differences are less dramatic than for the 
conduction-band states. The four and eight-band ground state energies agree 
to within $ 3~\rm meV$. For six bands, however, the ground state energy 
differs from the other two by $\approx 40~\rm meV$.  At first sight this is 
surprising, since one generally expects
the more complicated model to produce more accurate results. 
However,  for InAs $\Delta \approx E_g$ so 
 if  either  the conduction or split-off band is to be
included, then both should be.
The six-band model violates this requirement by allowing mixing of the
valence-band states, but leaving the conduction band decoupled.
The eight and four-band models do predict slightly 
different level spacings, but the basic pattern is the same. $E_1 - E_0$ is
large ( $20~\rm meV$ for four bands, $16~\rm meV$ for eight bands) with
a smaller spacing between excited states ($2~\rm meV$ and $8 ~\rm meV$ for
four and eight bands respectively).

An interesting way of viewing the model dependence of the energy
is to compare with the
size dependence. As a simple example, comparison of Fig.'s 4a and 7a 
shows that a $b=14 ~\rm nm$
island calculated with $m(\vec r )$ gives the same ground state
energy as the eight-band model at $b=12.5 ~\rm nm$. Hence, if the uncertainty
in the island geometry is greater than $1.5 ~\rm nm$ we would expect these
inaccuracies to dominate the errors in the electronic structure results.
This gives an indication of the importance of specifying the correct
island geometry.

\section{Conclusions}

Coupling the valence and conduction bands has a strong
 impact on the spectrum of  InAs quantum dot states.
The eight-band model gives results significantly different 
than one, four, and six-band approximations.
It predicts larger binding energies
and strongly confined excited states which do not appear in 
one-band  approximations.

The results presented here clearly demonstrate the need
for eight-band calculations, or perhaps even more complex techniques such
as pseudopotentials.\cite{fu} 
While large scale calculations using complex Hamiltonians have become
feasible, the results are no better than the input parameters used. 
Accurate agreement between theory and experiment will require
precise measurements of the island geometry.

\phantom{a}

I wish to thank Mats-Erik Pistol, Mark Miller, and  Jonas Ohlsson
for enlightening comments and discussions.


\begin{references}
\bibitem[*]{email}  e-mail: cpryor@zariski.ftf.lth.se
\bibitem{bastard} J. -Y. Marzin and G. Bastard, Solid State Commun. 
{\bf 91}, 39 (1994).

\bibitem{grundmann}  M. Grundmann, O. Stier, and D. Bimberg, 
 Phys. Rev. B {\bf 52},  11969 (1995).

\bibitem{cusack} M. A. Cusack, P. R. Briddon, and M. Jaros, 
Phys. Rev. B  {\bf 54}, 39 (1996).

\bibitem{pryor} C. Pryor, M-E. Pistol, L. Samuelson, 
http://xxx.lanl.gov/abs/cond-mat/9705291

\bibitem{fu} H. Fu, A. Zunger, 
Phys. Rev. B  {\bf 55}, 1642 (1997). 

\bibitem{stier} O. Stier, D. Bimberg,
Phys. Rev. B  {\bf 55}, 7726 (1997).

\bibitem{LL}   L.D. Landau and E.M. Lifshitz, {\it Theory of elasticity},
(Pergamon, London, 1959).

\bibitem{bahder} T.  B. Bahder, 
Phys. Rev. B  {\bf 45}, 1629 (1992);
 T.  B. Bahder, 
Phys. Rev. B  {\bf 41}, 11992 (1990);

\bibitem{LB} {\it Semiconductors}, edited by O. Madelung, 
M. Schilz, and H. Weiss, 
Landolt-B\"{o}rnstein, Numerical Data and Functional Relations in Science 
and Technology Vol. 17 (Springer-Verlag, New York, 1982).

\bibitem{DP} D. D. Nolte, W. Walukiewicz, and E. E. Haller, 
Phys. Rev. Lett.  {\bf 59}, 501 (1987);

\bibitem{offset} S. Tiwari and D. J. Frank, 
Appl. Phys. Lett.  {\bf 60}, 630 (1992).

\bibitem{e14} S. Adachi, 
J. Appl. Phys. {\bf 53}, 8775 (1982).
\end{references}
\end{document}